\documentclass{iau}
\usepackage{graphicx}

\title[Dust emission in ETG with HeViCS] 
{Dust Emission in Early-Type Galaxies\\ with the Herschel Virgo Cluster Survey}

\author[Sperello di Serego Alighieri]   
{Sperello di Serego Alighieri$^1$ \& members of the HeViCS team
 }

\affiliation{$^1$INAF - Osservatorio Astrofisico di Arcetri,\\
Largo E. Fermi 5, 50125 Firenze, Italy \\ email: {\tt sperello@arcetri.astro.it} \\[\affilskip]
}

\pubyear{2013}
\volume{295}  
\pagerange{xx--xx}
\jname{The intriguing life of massive galaxies}
\editors{D. Thomas, A. Pasquali \& I. Ferreras, eds.}
\begin{document}

\maketitle

\begin{abstract}
We have searched for dust in an optical sample of 910 Early-Type Galaxies (ETG) in the Virgo cluster
(447 of which are optically complete at $m_{pg} \leq 18.0$),
extending also to the dwarf ETG, using \textit{Herschel} images at 100, 160, 250, 350 and 500 $\mu$m. Dust
was found in 52 ETG (46 are in the optically complete sample), including M87 and another 3 ETG with strong
synchrotron emisssion. Dust is detected in 17\% of ellipticals, 41\% of lenticulars,
and in about 4\% of 
dwarf ETG. The dust-to-stars mass ratio increases with decreasing optical luminosity, and for some dwarf ETG reaches values similar to those of the 
dusty late-type galaxies. Slowly rotating ETG are more likely to contain dust than fast rotating ones.
Only 8 ETG have both dust and HI, while 39 have only dust and 8 have only HI, surprisingly showing that only rarely dust and HI survive together. ETG with dust appear to be concentrated in 
the densest regions of the cluster, while those with HI tend to be at the periphery. ETG with an X-ray active SMBH are more likely to have dust and vice versa the dusty ETG are more likely to have an active SMBH.
\end{abstract}

\firstsection 
\section{Introduction}

Most of the baryons in a cluster of galaxies are in the hot intracluster medium, some of which is associated with the most massive galaxies. The hot gas interacts in many ways with the cold phases of the interstellar medium of these galaxies, and these interactions have a fundamental effect on the evolution of the galaxies themselves. In order to understand the evolution of massive galaxies (the topic of this Symposium), particularly in clusters, it is therefore important to study also the coldest phases of their interstellar medium. The most massive galaxies are Early-Type Galaxies (ETG), and, since they may have formed by merging or accretion of smaller ones, it is useful to include in the study also the dwarf ETG. 
In \cite[di Serego Alighieri et al. (2007)]{diS07} we have systematically studied the neutral atomic gas (HI) content of a large and complete sample of ETG in the Virgo cluster, using the Arecibo Legacy Fast ALFA 21-cm survey \cite[(Giovanelli et al. 2007)]{Gio1}. HI is found in very few massive ETG, where the cold gas could have a recent external origin, and in a few peculiar dwarf galaxies at the edge of the ETG classification.
The \textit{Herschel Space Observatory} \cite[(Pilbratt et al. 2010)]{Pil10} is giving us the opportunity to study the dust content of the same sample of Virgo ETG. We have done so within the Herschel Virgo Cluster Survey \cite[(HeViCS, Davies et al. 2010)]{Dav10}, an open-time Key Programme for a confusion-limited imaging survey of a large fraction of the Virgo cluster in 5 bands: at 250, 350 and 500 $\mu$m with SPIRE \cite[(Griffin et al. 2010)]{Gri10} and at 100 and 160 $\mu$m with PACS \cite[(Poglitsch et al. 2010)]{Pog10}. We describe here the main results of this work. A more detailed and complete account has been submitted to A\&A (hereafter dSA12).

\section{Input sample, analysis and results}

We start with a sample of Virgo ETG selected in the optical from the GOLDMine compilation \cite[(Gavazzi et al. 2003)]{Gav03}, mostly based on the Virgo Cluster Catalogue \cite[(VCC, Binggeli et al. 1985)]{Bin85}, to be ETG (i.e. equal to or earlier than S0a) and excluding those with $v_{hel}<3000$ km/s. With these selection criteria 925 ETG are within the 4 HeViCS fields and constitute our input sample. Out of these, 447 are brighter than the VCC completeness limit ($m_{pg} \leq 18.0$) and form the optically complete part of our input sample. Out of the input sample, 287 ETG have inaccurate positions in the literature, based only on the original work of Binggeli et al. (1985), insufficient to find reliable counterparts in the HeViCS images. Using r-band SDSS images, we then remeasured the position for these ETG, except for 15 (all with $m_{pg} > 19.0$), for which the identification is unsure.
We have looked for a reliable far-IR counterpart in the HeViCS 250 $\mu$m mosaic image for all the 910 Virgo ETG with accurate coordinates, and found one for 52 of them at $S/N > 6$. For these sources we measured the flux in each of the 5 HeViCS bands using an aperture of 30 arcsec radius, large enough to contain the PSF also at 500 $\mu$m. For 12 sources, which have far-IR emission exceeding this aperture, we used larger apertures, up to 78 arcsec radius. 46 far-IR counterparts have $F_{250} \geq 25.4$ mJy, which is our completeness limit at 250 $\mu$m. 
We detect dust above the synchrotron component in the 4 ETG with radio emission, including M87. Given the large number of background sources present in the 250 $\mu$m images, following the methods of \cite[Smith et al. (2011)]{Smi11}, we estimate that on average 1.5 ETG (most likely dwarfs), out of our input sample of 910, have a far-IR counterpart which is a background source. We have used the distance given in GOLDMine, which distinguishes various components in the Virgo cluster at 17, 23 and 32 Mpc \cite[(Gavazzi et al. 1999)]{Gav99}.

Dust appears to be very concentrated, much more than stars. The only ETG with a considerable amout of off-nuclear dust is M86, where it appears to be mostly in a filament at 2 arcmin (about 10 kpc in projection) to the South-East \cite[(Gomez et al. 2010)]{Gom10}.
Dust masses and temperatures have been estimated for the 52 ETG with a far-IR counterpart by fitting a modified black-body to the measured far-IR fluxes, assuming a spectral index $\beta = 2$ and a MW emissivity, and taking into account colour and aperture corrections. We have also estimated stellar masses with the methods of \cite[Zibetti et al. (2009)]{Zib09}, using the available optical/IR broad-band photometry. The dust temperature ranges between 15 and 30 K, and correlates with the stellar mass and with the B-band average surface brigthness within the effective radius (dSA12). The latter correlation suggests that most of the dust heating is due to radiation produced by stellar sources.

\section{Discussion}

\begin{figure}[b]
\begin{center}
\includegraphics[width=11.7cm]{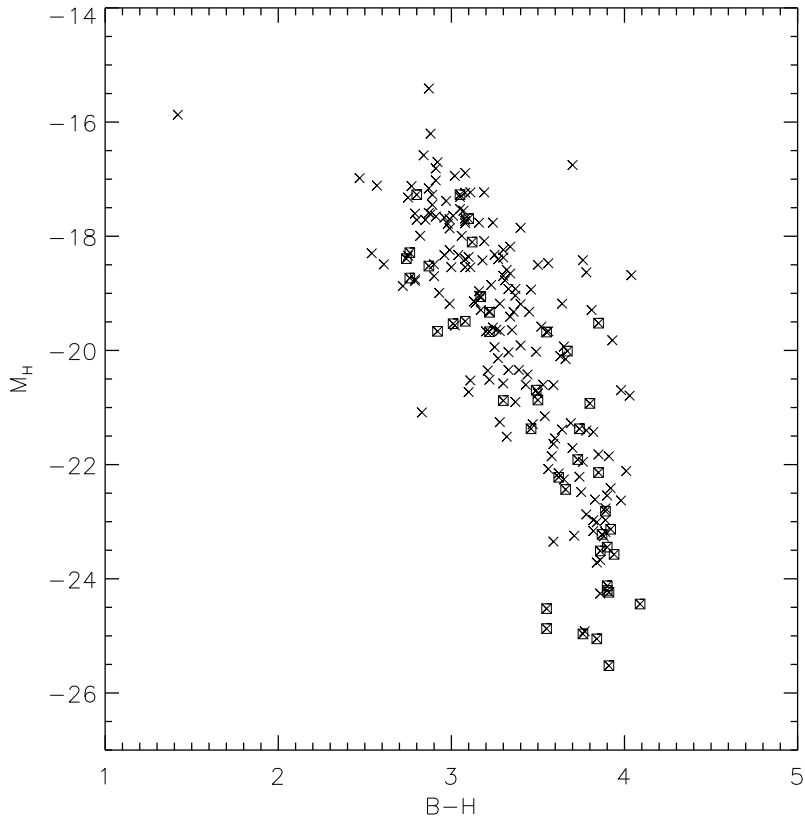} 
 \caption{Optical/IR CMD for the Virgo ETG with accurate photometry. The dusty ones (squares), and are not bluer, i.e. not more star-forming, than the other galaxies (crosses).}
   \label{fig1}
\end{center}
\end{figure}

Dust detection rates for the complete samples (i.e. 43 far-IR counterparts with $F_{250} \geq 25.4$ mJy out of the 447 input ETG with $m_{pg} \leq 18.0$ and accurate position) are 9.6\% for all ETG, 17.1\% for ellipticals, 41.4\% for lenticulars and 3.7\% for dwarf ETG. The latter rate becomes 3.6\%, if we take into account that 1.5 of the assumed far-IR counterpart of dwarf ETG are in fact counterparts of background sources (see the previous section), and that about 8 of the dwarf ETG of the input sample without a measured radial velocity are likely background galaxies. These rates are smaller than those previously measured on samples of bright ETG \cite[(Knapp et al. 1989, Temi et al. 2004, Smith et al. 2012)]{Smi12}, as can be expected since our sample extends to faint galaxies and the dust detection rate correlates strongly with optical luminosity (dSA12).
The dust-to-star mass ratio varies over almost 6 orders of magnitude, anticorrelates with the optical luminosity, and for some dwarf ETG reaches very high values (around a few $10^{-2}$), as high as for the dusty late-type galaxies. This is surprising, also given that the dusty ETG do not show signs of star formation. In fact the colours of the dusty ETG are not bluer than those of the non-dusty ones (Fig.\,\ref{fig1}).

The distinction between fast and slow rotators appears to be an important one for ETG \cite[(Emsellem et al. 2011, and references on the ATLAS$^{3D}$ project)]{Ems11}. For the ETG of our input sample the detailed kinematical information necessary for this distinction is available only for the 49 ones, which are in common with the ATLAS$^{3D}$ sample. Since we detect dust in $69\pm 23\%$ of the slow rotators (in 9 out of 13) and in $28\pm 9\%$ of the fast ones (in 10 out of 36), it appears that the former are considerably more likely to have dust. This is the opposite to what is seen for molecular gas in the whole of the 260 ETG of the ATLAS$^{3D}$ sample. In fact \cite[Young et al. (2011)]{You11} find that the CO detection rate is $6\pm 4\%$ in slow rotators and $24\pm 3\%$ in fast ones. This is surprising, since dust and molecular gas are thought to be closely associated \cite[(Draine et al. 2007, Corbelli et al. 2012)]{Dra07}; in fact, for the dust-detected ETG of our sample which have information on the molecular gas content, the dust-to-molecular-gas mass ratio is always $2 \times 10^{-2}$ within a factor of two, and lower limits are consistent with this range. We suggest that a possible explanation of this difference could be an environmental effect, since most of the ATLAS$^{3D}$ galaxies are outside of the Virgo cluster. In fact, of the 19 dust-detected ETG, which we have in common with the ATLAS$^{3D}$ sample, molecular gas is detected in 3 slow rotators and in 5 fast ones, a much more balanced situation than found by \cite[Young et al. (2011)]{You11} in the whole ATLAS$^{3D}$ sample, and all slow rotators with molecular gas in this whole sample are actually in the Virgo cluster.
The difference we find could be due to the presence of kinematically peculiar objects among the dusty slow rotators in the Virgo cluster, like galaxies with counter-rotating components mimicking slow rotation. We can exclude this possibility, since the brightest and most regular ellipticals and lenticulars in the Virgo cluster like M49, M84, M86, M87, M89, NGC 4261 and NGC 4526 are among the dusty slow rotators, reinforcing our suggestion about an environmental effect.

We have also looked at the relationship between dust and HI for the Virgo ETG, updating the work done by \cite[di Serego Alighieri et al. (2007)]{diS07} on the HI content of Virgo ETG to include the 4-8 degrees declination strip, which has become available in the ALFALFA HI survey since their work \cite[(Haynes et al. 2011)]{Hay11}. We find an intriguing incompatibility between dust and HI in Virgo ETG: we detect both dust and HI in only 8 ETG, while 39 ETG have dust but no HI, and 8 have HI but no dust. This dichotomy between dust and HI is reinforced by the position of the parent galaxies in the cluster. Dusty ETG appear to concentrate in the densest regions of the Virgo cluster, while HI-rich ETG tend to be at the periphery. While the presence of ETG with dust but no HI can be explained by the longer dust survival times and by the stronger effects of ram pressure stripping on HI, more difficult to understand is that there are ETG with HI but no dust, also given that these include two rather bright S0 galaxies: NGC 4262 and NGC 4270.

Concerning the relationship between the presence of dust and that of an AGN in Virgo ETG, our input sample has 71 ETG in common with the sample of Virgo ETG for which \cite[Gallo et al. (2010)]{Gal10} have looked for the presence of a supermassive black-hole (SMBH) by observing with \textit{Chandra} the nuclear X-ray luminosity down to a few $10^{38}$erg/s. Out of the 71 common ETG, 25 (35\%) have X-rays, most likely from a SMBH, and 14 (20\%) have dust. Of the X-ray luminous ETG 36\% have dust, and out of the dusty ETG 64\% are X-ray luminous. It appears that ETG with an X-ray active SMBH are more likely to have dust, and that ETG with dust are more likely to have an X-ray active SMBH.

\begin{thebibliography}{}

\bibitem[Bin85]{Bin85}{Binggeli, B., Sandage, A. \& Tammann, G.A., 1985, \textit{AJ}, 90, 1681}
\bibitem[Cor12]{Cor12}{Corbelli, E., Bianchi, S., Cortese, L., et al., 2012, \textit{A\&A}, 542, A32}
\bibitem[Dav10]{Dav10}{Davies, J.I., Baes, M., Bendo, G.J., et al., 2010, \textit{A\&A}, 518, L48}
\bibitem[diS07]{diS07}{di Serego Alighieri, S., Gavazzi, G., Giovanardi, C., et al., 2007, \textit{A\&A}, 474, 851}
\bibitem[Dra07]{Dra07}{Draine, B.T., Dale, D.A., Bendo, G., et al., 2007, \textit{ApJ}, 663, 866}
\bibitem[Ems11]{Ems11}{Emsellem, E., Cappellari, M., Krajnovi\'c, D., et al., 2011, \textit{MNRAS}, 414, 888}
\bibitem[Gal10]{Gal10}{Gallo, E., Treu, T., Marshall, P.J., et al., 2010, \textit{ApJ}, 714, 25}
\bibitem[Gav99]{Gav99}{Gavazzi, G., Boselli, A., Scodeggio, M., Pierini, D. \& Belsole, E., 1999, \textit{MNRAS}, 304, 595}
\bibitem[Gav03]{Gav03}{Gavazzi, G., Boselli, A., Donati, A., Franzetti, P. \& Scodeggio, M., 2003, \textit{A\&A}, 400, 451}
\bibitem[Gio2]{Gio2}{Giovanelli, R., Haynes, M.P., Kent, B.R., et al., 2007, \textit{AJ}, 133, 2569}
\bibitem[Gom10]{Gom10}{Gomez, H.L., Baes, M., Cortese, L., et al., 2010, \textit{A\&A}, 518, L45}
\bibitem[Hay11]{Hay11}{Haynes, M.P., Giovanelli, R., Martin, A.M., et al., 2011, \textit{AJ}, 142, 170}
\bibitem[Kna89]{Kna89}{Knapp, G.R., Guhathakurta, P., Kim, D.-W. \& Jura, M., 1989, \textit{ApJSS}, 70, 329}
\bibitem[Pil10]{Pil10}{Pilbratt, G.L., Riedinger, J.R., Passvogel, T., et al., 2010, \textit{A\&A}, 518, L1}
\bibitem[Smi11]{Smi11}{Smith, D.J.B., Dunne, L., Maddox, S.J., et al., 2011, \textit{MNRAS}, 416, 857}
\bibitem[Smi12]{Smi12}{Smith, M.W.L., Gomez, H.L., Eales, S.A., et al., 2012, \textit{ApJ}, 748, 123}
\bibitem[Tem04]{Tem04}{Temi, P., Brighenti, F., Mathews, W.G. \& Bregman, J.D., 2004, \textit{ApJSS}, 151, 237}
\bibitem[You11]{You11}{Young, L.M., Bureau, M., Davis, T.A., et al., 2011, \textit{MNRAS}, 414, 940}
\bibitem[Zib09]{GZib09}{Zibetti, S., Charlot, S. \& Rix, H.-W., 2009, \textit{MNRAS}, 400, 181}


\end{thebibliography}
\end{document}